\documentclass[a4paper,11pt,final,notitlepage]{article}
\usepackage{amsfonts}
\usepackage{amsmath}

\input{tcilatex}

\begin{document}

\title{The Role of Three-dimensional Subalgebra in the Analysis of Hidden
Symmetries of Differential Equations}
\author{Mladen Nikoli\'{c} and Milan Rajkovi\'{c} \\
%EndAName
Institute of Nuclear Sciences Vin\v{c}a, \\
P.O. Box 522, 11001 Belgrade, Serbia}
\maketitle

\begin{abstract}
Some new properties of symmetries that disappear as point symmetries after
the first reduction of order of an ODE and reappear after the second are
analyzed from the aspect of three-dimensional subalgebra of symmetries of
differential equations. The form of a hidden symmetry is shown to consist of
two parts, one of which always remains preserved as a point symmetry, and
the second (fundamental) part which behaves as the complete hidden symmetry.
Symmetry that disappears as point symmetry and remains hidden (non-local)
during $n$ reductions of order before reappearing as a point symmetry is
also introduced and termed convertible symmetry of order $n-1.$ We discuss
the necessity for such classification in order to distinguish them from
hidden symmetries of type I and type II, which are defined with respect to
reduction of order by one.
\end{abstract}

\section{Introduction}

The search for Lie point symmetries represents a common method for
determining the properties of ordinary differential equations (ODEs) so that
the found symmetries are used to reduce the order of the equation. If a
differential equation of order $n$ has an $n$-parameter solvable Lie group
of symmetries, the equation may be solved by group-theoretic quadrature
method following a proper reduction path using one-parameter subgroups. This
approach may not be possible due to, for example, non solvability of the $n$%
-parameter Lie group so that certain symmetries may be lost making the
complete reduction impossible. In that case, the lower order reduced ODE has
a type I hidden symmetry\cite{Barbara1},\cite{Barbara2},\cite{Barbara4},\cite%
{Barbara5} whose origin may be attributed to the reduction by the invariants
of a non-normal subgroup, i.e. a symmetry group that is lost is not a normal
subgroup. In the case of two point symmetries, $X$ and $Y$ let%
\begin{equation*}
\lbrack X,Y]=X,
\end{equation*}%
$X$ being the normal subgroup. Reducing the order of equation using the
normal subgroup $X$ leads to the descendant of $Y$ being a point symmetry of
the reduced equation. On the other hand, reduction using $Y$ first, causes $%
X $ to be lost as a point symmetry of the descendant equation. Complete
reduction of the starting equation of order $n,$ in general, may also be
impossible if the symmetry group is not an $n$-parameter group, i.e. if it
has less than $n$ parameters. Exceptions to this case arise when during
reduction new symmetries appear in addition to inherited ones when the order
of the ODE is reduced by one, and they are known as type II hidden symmetries%
\cite{Barbara1},\cite{Barbara2},\cite{Barbara3},\cite{Barbara4},\cite%
{Barbara5}. In this scenario a symmetry of the initial ODE may not be
inherited in the reduced equation (becomes nonlocal), however it transforms
to an inherited symmetry in the subsequent reduction. As it turns out, in
order to solve an ODE the requirement that the reduction of order must be by
the normal subgroup is not necessary\cite{Geronimi}. Moreover, the complete
reduction of the starting ODE is more likely by inclusion of a larger
variety of symmetries, and also the chances for appearance of type II hidden
symmetries are increased, enhancing the likelihood of complete reduction. In
the following sections we analyze the conditions necessary for a symmetry to
disappear during the first reduction and become non-local (type I hidden
symmetry), and to reappear as a point symmetry (type II hidden symmetry)
after the second reduction of order. We call this symmetry a convertible
type symmetry of order I. In general, a convertible type of symmetry of
order $n-1$ may be defined as a point symmetry that disappears during the
first reduction, remains hidden (non-local) during $n-1$ reductions, and
reappears as a point symmetry after $n$ reductions. Convertible symmetries
may be regarded as a special class of hidden symmetries of type II. We show
that such symmetries consist of two parts, one of which always remains of
point type and the other of convertible (changeable) type which has the
power over the behavior of the complete symmetry. By knowing the exact form
of the hidden symmetry of type II, as shown in this paper, one may infer
whether the origin of the hidden symmetry lies in convertible symmetry or
non-local symmetry of unknown, higher order than the order of the initial
ODE. The paper is organized as follows: In the first Section the focus is on
three generators of symmetries from the complete algebra of symmetries of
the $n$-th order ODE\ and the general form of a convertible symmetry of type
I, i.e. a hidden symmetry that transforms from type I to type II after two
consecutive reductions of order. In sections two and three, we discuss three
examples that employ results of the previous Section. The first example is
specially constructed in order to illuminate the results of Section $1$,
while the second example shows how performing a nonnormal reduction along
with the use of hidden symmetry properties provides an easier path to the
solution of the third order ODE. In the third example, presented in a
separate section, we analyze a convertible symmetry of type II (a symmetry
that remains hidden during two reductions of order and reappears as a point
symmetry after the third reduction) Moreover we show how important
information about the integration and reduction paths may be known in
advance by exploiting the results presented here and the properties of the
solvable algebra. Two appendices at the end of the paper complement the main
exposition. The first one contains a derivation of the practical rule
according to which it may be inferred whether a symmetry along the reduction
path is of point or non-local type. The second one illustrates an
alternative derivation of the properties demonstrated in Section $1$. In the
final Section we give some concluding remarks.

\section{Properties of Convertible Symmetry of Type I}

We consider the general $n$-th order ODE%
\begin{equation}
y^{(n)}=f_{n}(x,y,y^{\prime },...y^{(n-1)}),  \label{1}
\end{equation}%
where $x$ and $y$ are the independent and dependent variables, respectively,
and $y^{(n)}=dy^{n}/dx^{n},$ admitting three vector fields of the
corresponding symmetries $X,$ $Y$ and $Z.$ Naturally, eq. (\ref{1}) may have
more symmetry generators, hence $X,$ $Y$ and $Z$ may not necessarily form an
algebra. In accordance with the properties of the hidden symmetries
mentioned above,\ let%
\begin{equation*}
\lbrack X,Y]=\lambda X,
\end{equation*}%
where $\lambda $ is a constant (either 0 or scaled to 1), be the condition
for preservation of symmetry $Y$ when the reduction is performed using
symmetry $X$. In addition, let%
\begin{equation*}
\lbrack X,Z]\neq \mu X
\end{equation*}%
represent the condition for the symmetry $Z$ to disappear when the reduction
is performed using $X$ ($\mu $ is a constant). Then during reduction using
Lie point symmetry $X$ the order of the initial ODE is reduced by one and $Y$
is inherited as a point symmetry, while $Z$ becomes non-local symmetry of
the new differential equation, i.e. $Z$ is lost as a point symmetry. The
canonical form of 2-dimensional Lie algebras\cite{Lie}, representing four
distinct classes of algebras equivalent under point transformations may be
described simply in terms of the commutator and the pseudo-scalar product%
\begin{equation*}
X\vee Y=\xi _{1}\eta _{2}-\xi _{2}\eta _{1},
\end{equation*}%
where%
\begin{equation*}
X=\xi _{1}(x,y)\frac{\partial }{\partial x}+\eta _{1}(x,y)\frac{\partial }{%
\partial y},\text{ and }Y=\xi _{2}(x,y)\frac{\partial }{\partial x}+\eta
_{2}(x,y)\frac{\partial }{\partial y}.
\end{equation*}%
The canonical form includes four algebras, two Abelian and two solvable,
whose properties are given in Table 1. 
\begin{equation*}
\begin{tabular}{|l|l|l|l|l|}
\hline
Type & \multicolumn{2}{|l|}{$L_{2}$ structure} & \multicolumn{2}{|l|}{Basis
of $L_{2}$ in canonical variables} \\ \hline
I & $[X,Y]=0$ & $X\vee Y=0$ & $X=\frac{\partial }{\partial s}$ & $Y=r\frac{%
\partial }{\partial s}$ \\ \hline
II & $[X,Y]=0$ & $X\vee Y\neq 0$ & $X=\frac{\partial }{\partial s}$ & $Y=%
\frac{\partial }{\partial r}$ \\ \hline
III & $[X,Y]=X$ & $X\vee Y=0$ & $X=\frac{\partial }{\partial s}$ & $Y=s\frac{%
\partial }{\partial s}$ \\ \hline
IV & $[X,Y]=X$ & $X\vee Y\neq 0$ & $X=\frac{\partial }{\partial s}$ & $Y=r%
\frac{\partial }{\partial r}+s\frac{\partial }{\partial s}$ \\ \hline
\end{tabular}%
\end{equation*}

Let $(r,s)$ be canonical coordinates for the group generated by the
infinitesimal generator of a one-parameter Lie group $X$, so that%
\begin{equation*}
X=\frac{\partial }{\partial s}.
\end{equation*}%
The ODE (\ref{1}) written in terms of canonical coordinates is of the form%
\begin{equation*}
s^{(n)}=\Omega _{n}(r,s^{\prime },...,s^{(n-1)}),\text{ \ \ \ }s^{(k)}=\frac{%
d^{k}s}{dr^{k}}.
\end{equation*}%
Making a substitution $v=s^{\prime }$ we reduce the original equation to an
ODE of order $(n-1):$%
\begin{equation}
v^{(n-1)}=\Omega _{n-1}(r,v,...,v^{(n-2)}),\text{ \ \ }v^{(k)}=\frac{d^{k+1}s%
}{dr^{k+1}}.  \label{2}
\end{equation}%
Assuming that the third symmetry $Z$ may be expressed in canonical
coordinates as%
\begin{equation*}
Z=A(r,s)\partial _{r}+B(r,s)\partial _{s}
\end{equation*}%
we explore all four possibilities:

\bigskip

\begin{enumerate}
\item 
\begin{tabular}{|l|}
\hline
$X=\partial _{s}$, $\ \ \ Y=r\partial _{s},$ $\ \ \ \ Z=A(r,s)\partial
_{r}+B(r,s)\partial _{s}$ \\ \hline
\end{tabular}

The first prolongations of vector fields $Y$ and $Z$ are:%
\begin{eqnarray*}
Y^{(1)} &=&r\partial _{s}+\partial _{v}, \\
Z^{(1)} &=&A\partial _{r}+B\partial
_{s}+(B_{r}+(B_{s}-A_{r})v-A_{s}v^{2})\partial _{v}.
\end{eqnarray*}%
ODE (\ref{2}) has an inherited point symmetry from eq. (\ref{1})%
\begin{equation*}
\tilde{Y}^{(1)}=\partial _{v},
\end{equation*}%
where the tilde sign denotes restriction of the inherited generator to the
fundamental differential invariants of one local group generator. The
prolongation $Z^{(1)}$ for arbitrary $A$ and $B$ is a nonlocal symmetry.
Hence, ODE (\ref{2}) may be further reduced using vector field $Y^{(1)}.$
Since $\tilde{Y}^{(1)}=\partial _{v}$ is the symmetry of eq. (\ref{2}), it
follows that%
\begin{equation*}
\frac{\partial \Omega _{n-1}}{\partial v}=0,
\end{equation*}%
i.e.%
\begin{equation}
v^{(n-1)}=\Omega _{n-1}(r,v^{\prime },...,v^{(n-2)}).  \label{3}
\end{equation}%
Introducing new independent variable%
\begin{equation*}
w=\frac{dv}{dr}
\end{equation*}%
eq. (\ref{3}) may be written as:%
\begin{equation}
w^{(n-2)}=\Omega _{n-2}(r,w,...,w^{(n-3)})  \label{4}
\end{equation}%
The second prolongation of the vector field $Z$ is:%
\begin{eqnarray*}
Z^{(2)} &=&A\partial _{r}+B\partial
_{s}+(B_{r}+(B_{s}-A_{r})v-A_{s}v^{2})\partial _{v}+ \\
&&[B_{rr}+(2B_{rs}-A_{rr})v+(B_{ss}-2A_{rs})v^{2}-A_{ss}v^{3}+ \\
&&(B_{s}-2A_{r}-3A_{s}v)w]\partial _{w},
\end{eqnarray*}%
so that if $Z^{(2)}$ is reduced into a point symmetry of ODE (\ref{4}), then:%
\begin{equation*}
\tilde{Z}^{(2)}=\alpha (r,w)\partial _{r}+\beta (r,w)\partial _{w}.
\end{equation*}%
Furthermore%
\begin{equation*}
Z^{(2)}r=A=\alpha (r)\text{ implying }A_{s}=0\text{ so that }A=A(r),
\end{equation*}%
and%
\begin{equation*}
Z^{(2)}w=B_{rr}+(2B_{rs}-A^{\prime \prime
}(r))v+B_{ss}v^{2}+(B_{s}-2A^{\prime }(r))w=\beta (r,w).
\end{equation*}%
Equating terms of equal powers of $v$ and $w$, in a straightforward manner
one obtains%
\begin{eqnarray*}
A &=&c_{1}r^{2}+c_{3}r+c_{4}, \\
B &=&(c_{1}r+c_{2})s+b(r),
\end{eqnarray*}%
where $c_{i}(i=1,2,3,4)$ are arbitrary constants, so that%
\begin{equation}
Z=(c_{1}r^{2}+c_{3}r+c_{4})\partial _{r}+[(c_{1}r+c_{2})s+b(r)]\partial _{s}
\label{5}
\end{equation}%
and%
\begin{equation}
\tilde{Z}^{(2)}=(c_{1}r^{2}+c_{3}r+c_{4})\partial _{r}+[b^{\prime \prime
}(r)+(c_{2}-2c_{3}-3c_{1}r)w]\partial _{w.}  \label{6}
\end{equation}%
For $c_{1}\neq 0,$ the vector field $Z$ disappears during first reduction
and reappears after two reductions as a point symmetry \ (\ref{6}). For $%
c_{1}=0,$ $[X,Z]=$ $c_{2}X$ holds, and $Z$ is preserved as a point symmetry
during first reduction. Hence, this case requires no further consideration.
If the vector field in eq. (\ref{5}) is written as%
\begin{equation}
Z=Z^{0}+Z^{\ast },  \label{7}
\end{equation}%
where%
\begin{equation}
Z^{0}=(c_{3}r+c_{4})\partial _{r}+(c_{2}s+b(r))\partial _{s},  \label{8}
\end{equation}%
and%
\begin{equation}
Z^{\ast }=c_{1}(r^{2}\partial _{r}+rs\partial _{s}),  \label{9}
\end{equation}%
it may be shown, that%
\begin{eqnarray}
\lbrack X,Z^{\ast }] &\neq &X,\text{ \ }[X,Z^{0}]=X,  \label{10} \\
\lbrack Y,Z^{\ast }] &\neq &Y,\text{ \ }[Y,Z^{0}]=Y.  \notag
\end{eqnarray}%
The same relationships hold if $X$ and $Y$ are replaced with $X^{(1)}$ and $%
Y^{(1)}$ respectively. Therefore, the vector field $Z^{0}$ represents the
unalterable (fundamental) symmetry in the sense that it is preserved as a
point symmetry during both reductions (irrespective of the first symmetry
used for reduction), while $Z^{\ast }$ represents the dominant symmetry
which is lost during the first reduction and reappears after the second
reduction, and whose behavior determines the behavior of the symmetry $Z$.
Hence we may use expression (\ref{7}) as a general form of the type I hidden
symmetry that turns into a type II hidden symmetry. In evaluating the last
commutator relation in (\ref{10}) we have used the fact that canonical
coordinates are not uniquely defined and that they satisfy transformation%
\begin{eqnarray}
\bar{s} &=&s+G(r)  \label{10a} \\
\bar{r} &=&F(r),  \notag
\end{eqnarray}%
for arbitrary smooth functions $F$ and $G$, with additional constraint $%
F^{^{\prime }}(r)\neq 0.$ Furthermore, this nonuniqueness property of
canonical coordinates allows the following general representation:%
\begin{eqnarray}
Z^{0} &=&(\alpha r+\beta )\partial _{r}+\gamma s\partial _{s},  \label{11} \\
Z^{\ast } &=&r^{2}\partial _{r}+rs\partial _{s},  \notag
\end{eqnarray}%
where $\alpha ,$ $\beta $ and $\gamma $ are constants (equal to 0 or scaled
to 1).

\item 
\begin{tabular}{|l|}
\hline
$X=\partial _{s}$, $\ \ \ Y=\partial _{r},$ $\ \ \ \ Z=A(r,s)\partial
_{r}+B(r,s)\partial _{s}$ \\ \hline
\end{tabular}

During first reduction using vector field $X=\partial _{s}$ one gets in new
coordinates $(r,v=ds/dr):$%
\begin{equation*}
\tilde{Y}^{(1)}=\partial _{r}.
\end{equation*}%
If new canonical coordinates $(\rho ,\vartheta )$ are used so that%
\begin{equation*}
\tilde{Y}^{(1)}=\partial _{\vartheta },
\end{equation*}%
an interchange of independent and dependent variables yields 
\begin{equation*}
\frac{d\vartheta ^{n-1}}{d\rho ^{n-1}}=\Omega _{n-1}(\rho ,\vartheta
^{\prime },...,\vartheta ^{(n-2)}),
\end{equation*}%
and introducing 
\begin{equation*}
w=\frac{d\vartheta }{d\rho }=\frac{1}{v^{\prime }},
\end{equation*}%
one gets%
\begin{equation*}
w^{(n-2)}=\Omega _{n-1}(\rho ,w,...,w^{(n-3)}).
\end{equation*}%
The restriction of the second prolongation $Z^{(2)}$ in new coordinates $%
(\rho ,w)$ gives:%
\begin{gather*}
Z^{(2)}\rho =A_{r}+(B_{s}-B_{r})\rho -A_{s}\rho ^{2}=\alpha (\rho ,w), \\
Z^{(2)}w=[B_{rr}+(2B_{rs}-A_{rr})\rho +(B_{ss}-2A_{rs})\rho ^{2} \\
-A_{ss}\rho ^{3}+(B_{s}-2A_{r}-3A_{s}\rho )\frac{1}{w}](-w^{2})=\beta (\rho
,w).
\end{gather*}%
Equating terms of equal powers of $v$ and $w$, one obtains in a
straightforward manner%
\begin{equation*}
Z=(c_{1}s+c_{2}r+c_{3})\partial _{r}+((c_{1}-c_{4})r+c_{5}s+c_{6})\partial
_{s},
\end{equation*}%
Linear combinations with $X$ and $Y$ and non-uniqueness of canonical
coordinates yield%
\begin{equation*}
Z=s\partial _{r}+\alpha r\partial _{r}+\beta r\partial _{s}+\gamma s\partial
_{s},
\end{equation*}%
where $\alpha ,$ $\beta $ and $\gamma $ are constants. As in Case $1$, $Z$
may be written as the sum of $Z^{0}$ (preserved symmetry) and $Z^{\ast }$
(dominating, lost symmetry), with%
\begin{eqnarray*}
Z^{\ast } &=&s\partial _{r}, \\
Z^{0} &=&\alpha r\partial _{r}+\beta r\partial _{s}+\gamma s\partial _{s}.
\end{eqnarray*}

\item 
\begin{tabular}{|l|}
\hline
$X=\partial _{s}$, $\ \ \ Y=r\partial _{r},$ $\ \ \ \ Z=A(r,s)\partial
_{r}+B(r,s)\partial _{s}$ \\ \hline
\end{tabular}

Applying the same procedure as in previous two cases, one obtains 
\begin{equation*}
Z=Z^{0}=(\alpha r+\beta )\partial _{s},\text{ \ \ \ \ \ \ \ \ \ }Z^{\ast }=0.
\end{equation*}%
The vector field $Z$ is preserved as a point symmetry during the first
reduction, so in this case $Z$ is not a hidden symmetry.

\item 
\begin{tabular}{|l|}
\hline
$X=\partial _{s}$, $\ \ \ Y=s\partial _{s},$ $\ \ \ \ Z=A(r,s)\partial
_{r}+B(r,s)\partial _{s}$ \\ \hline
\end{tabular}

In this case the vector field $Z$ assumes the form:%
\begin{equation*}
Z=Z^{0}=\alpha \partial _{r}+(\beta r+\gamma )\partial _{s},\text{ \ \ \ \ \
\ \ \ \ }Z^{\ast }=0.
\end{equation*}%
and this symmetry is preserved during the first reduction as in Case $3$, so
again $Z$ is not a hidden symmetry.
\end{enumerate}

Denoting by $Z_{A}$ and $Z_{B}$ symmetries corresponding to Cases $1$ and $2$
respectively, i.e.%
\begin{eqnarray*}
Z_{A} &=&(r^{2}+(\gamma -\alpha )-\beta )\partial _{r}+(r+\gamma )s\partial
_{s}, \\
Z_{B} &=&(s+\alpha r)\partial _{r}+(\beta r+\gamma s)\partial _{s},
\end{eqnarray*}%
and remembering that in both cases $X$ and $Y$ commute, the following set of
relations is obtained%
\begin{eqnarray}
\lbrack X,Y] &=&0,  \label{100} \\
\lbrack X,Z_{i}] &=&\gamma X+Y,\text{ \ \ \ \ \ }i=A,B,  \notag \\
\lbrack Y,Z_{i}] &=&\beta X+\alpha Y,\text{ \ \ \ }i=A,B.  \notag
\end{eqnarray}%
In order to shed more light on the properties of hidden symmetries, in
Appendix B we present an alternative derivation of the above conditions
using algebraic properties of the subalgebra $Span(X,Y,Z)$. Hence, if a
hidden symmetry of type $Z_{A}$ or $Z_{B}$ occurs, then within the complete
symmetry group there is a three parameter symmetry group with vector fields%
\begin{equation*}
X=\partial _{s},\ \ Y=r\partial _{s},\ \ Z_{A}
\end{equation*}%
or%
\begin{equation*}
X=\partial _{s},\ \ Y=\partial _{r},\ \ Z_{B},
\end{equation*}%
respectively. Moreover, eq. (\ref{100}) indicates that vector fields $X,Y$
and $Z$ necessarily form a subalgebra $\mathcal{L}$ which may be a complete
algebra of initial ODE symmetries. Appropriate choice of the basis along
with the suitable choice of parameters $\alpha ,\beta $ and $\gamma $ may
cast, for example, the commutator relations of algebra $\mathcal{L=}$ $%
Span(X,Y,Z).$ in the form%
\begin{equation}
\lbrack \bar{X},Y]=0,\text{ \ \ \ \ \ }[\bar{X},Z]=\bar{X},\text{ \ \ \ \ \ }%
[Y,Z]=\bar{X},  \label{1.13}
\end{equation}%
where the choice was $\alpha =\beta =\gamma =1$ and $\bar{X}=X+Y.$ Another
choice of the basis may yield, using the nomenclature of ref. \cite{Mubarak}%
, two algebras%
\begin{eqnarray}
A_{3,6} &:&\text{ \ \ \ \ }[X,Z]=-Y,\text{ \ \ \ \ \ \ \ \ \ }[Y,Z]=X
\label{1.14} \\
A_{3,7} &:&\text{ \ \ \ \ }[X,Z]=lX-Y,\text{ \ \ \ \ }[Y,Z]=X+lY,\text{ \ \
\ \ }(l>0).  \notag
\end{eqnarray}%
In the case of algebra in (\ref{1.13}), standard reduction procedure using $%
\bar{X}$ may lead to solution by quadratures, however an example presented
in the next Section shows how nonnormal reduction starting with symmetry $Y$
leads to a less complicated solution procedure due to the properties of the
hidden symmetry $Z$.

\section{Applications of convertible symmetry properties}

\subsection{Solution of a third-order ODE}

For this example we use the three dimensional solvable algebra $A_{3,6}$:%
\begin{equation*}
X=\partial _{s},\text{ \ \ \ \ }Y=r\partial _{s},\text{ \ \ \ \ }%
Z=(r^{2}+1)\partial _{r}+rs\partial _{s}
\end{equation*}

with

\begin{equation*}
\lbrack X,Y]=0,\text{ \ \ \ \ }[X,Z]=Y,\text{ \ \ \ \ }[Y,Z]=-X.
\end{equation*}%
One of the forms of the most general third-order ODE with these symmetries is%
\begin{equation*}
F(s^{\prime \prime 2/3}(1+r^{2}),s^{\prime \prime -5/3}s^{\prime \prime
\prime }+\frac{3r}{1+r^{2}}s^{\prime \prime -2/3})=0,
\end{equation*}%
which upon selecting a specific class of equations yields%
\begin{equation*}
s^{\prime \prime \prime }+\frac{3r}{1+r^{2}}s^{\prime \prime }=s^{\prime
\prime 5/3}f(s^{\prime \prime 2/3}(1+r^{2})).
\end{equation*}%
Reduction using $X=\partial _{s}$ and introducing $v=ds/dr$ \ lowers the
order of the ODE:%
\begin{equation*}
v^{\prime \prime }+\frac{3r}{1+r^{2}}v^{\prime }=v^{5/3}f(v^{\prime
2/3}(1+r^{2}).
\end{equation*}%
The vector field $Y^{(1)}$ reduces to $Y^{(1)}=\partial _{v}$ in $(r,v)$
coordinates and hence remains a point symmetry as a consequence of the
commutator relationship $[X,Y]=0.$ On the other hand, the vector field $Z$
is lost as a point symmetry due to the commutator relationship $[X,Z]=Y.$
Note that according to the categorization of Section $1$, the symmetry $Z$
is of type $A$. Further reduction is therefore possible so that using $%
\tilde{Y}^{(1)}$ and introducing $w=dv/dr$ one obtains%
\begin{equation}
w^{\prime }+\frac{3r}{1+r^{2}}w=w^{5/3}f(w^{2/3}(1+r^{2})),  \label{2.6}
\end{equation}%
Due to the specific construction of the vector field $Z$ and on the basis of
relationship (\ref{6}), $Z^{(2)}$ reduces to point symmetry in coordinates $%
(r,w)$:%
\begin{equation*}
\tilde{Z}^{(2)}=(r^{2}+1)\partial _{r}-3rw\partial _{w},
\end{equation*}%
and makes the reduction to quadratures feasible. It should be remarked that
the reduction path chosen distinguishes out a class of first-order ODEs
solvable due to the symmetry $\tilde{Z}^{(2)}.$ Within this class, for
example, there is an ODE of Riccati type%
\begin{equation*}
w^{\prime }+\frac{3r}{1+r^{2}}w=(1+r^{2})^{-5/2}+(1+r^{2})^{1/2}w^{2}.
\end{equation*}%
Equation (\ref{2.6}) reduces to quadratures with the introduction of
canonical coordinates $(\rho ,\vartheta )$, so that%
\begin{equation*}
\frac{dr}{1+r^{2}}=\frac{dw}{-3rw}=\frac{d\vartheta }{1},
\end{equation*}%
yielding%
\begin{equation*}
\rho =w^{2/3}(1+r^{2}),\text{ \ \ \ \ }\vartheta =\arctan r.
\end{equation*}%
Hence, (\ref{2.6}) transforms into:%
\begin{equation*}
\frac{d\vartheta }{d\rho }=\frac{3}{2\rho ^{2}f(\rho )},
\end{equation*}%
finally generating%
\begin{equation*}
\vartheta =\frac{3}{2}\int \frac{d\rho }{\rho ^{2}f(\rho )}+const.
\end{equation*}

\subsection{Nonnormal reduction of order using convertible symmetry}

Reduction of order of an ODE possessing a solvable Lie algebra of symmetries
is a well defined method, usually performed by means of the normal subgroup.
However, nonnormal reduction may prove to offer a more efficient and easier
route to the solution if properties of hidden (convertible) symmetries are
used, as illustrated in the next example. Consider the third order ODE:%
\begin{equation}
y^{\prime \prime \prime }+\frac{3}{x}y^{\prime \prime }+xy^{\prime \prime
2}=0.  \label{13}
\end{equation}%
The equation possesses the solvable Lie algebra of symmetries%
\begin{equation*}
X=\partial _{y}\text{ \ \ }Y=x\partial _{y}\text{ \ \ }Z=x^{2}\partial
_{x}+xy\partial _{y},
\end{equation*}%
with commutator relations%
\begin{equation*}
\lbrack X,Y]=0,\text{ \ \ }[X,Z]=Y\text{, \ \ }[Y,Z]=0.
\end{equation*}%
The standard procedure would be to perform the order reduction using the
normal group $Y=x\partial _{y},$ yielding the canonical coordinates $r=x$
and $s=y/x.$ Introducing $v=ds/dr,$ initial equation is reduced to a second
order ODE 
\begin{equation}
\frac{d^{2}v}{dr^{2}}+(\frac{6}{r}+4rv)\frac{dv}{dr}+\frac{6}{r^{2}}%
v+4v^{2}+r^{2}\left( \frac{dv}{dr}\right) ^{2}=0.  \label{14}
\end{equation}%
This equation has at least two point symmetries $\tilde{X}^{(1)}$ and $%
\tilde{Z}^{(1)}$ so the complete reduction is possible, however further
calculations are complicated. On the other hand nonormal reduction using $%
X=\partial _{y},$ yields%
\begin{equation}
\frac{d^{2}v}{dr^{2}}+\frac{3}{r}\frac{dv}{dr}+r\left( \frac{dv}{dr}\right)
^{2}=0.  \label{15}
\end{equation}%
Reductions of vector fields $Y$ and $Z$ in canonical coordinates yield%
\begin{eqnarray*}
\tilde{Y}^{(1)} &=&\partial _{v} \\
\tilde{Z}^{(1)} &=&r^{2}\partial _{r}+(\dint vdr-rv)\partial _{v}.
\end{eqnarray*}%
Hence, $\tilde{Y}^{(1)}$ is a point symmetry, and as it turns out the only
point symmetry of eq. (\ref{15}), while $\tilde{Z}^{(1)}$ is clearly a
nonlocal symmetry. Note that symmetry $Z$ satisfies relationship (\ref{100}%
), with $\gamma =0$ Further reduction using $\tilde{Y}^{(1)}$and introducing 
$w=dv/dr$ yields the Bernoulli equation%
\begin{equation}
\frac{dw}{dr}+\frac{3}{r}w+rw^{2}=0.  \label{16}
\end{equation}%
with point symmetry%
\begin{equation*}
\tilde{Z}^{(2)}=r^{2}\partial _{r}-3rw\partial _{w}.
\end{equation*}%
Using new canonical coordinates%
\begin{eqnarray*}
\rho  &=&r^{3}w, \\
\vartheta  &=&-\frac{1}{r}
\end{eqnarray*}%
eq. (\ref{16}) is transformed to%
\begin{equation*}
\frac{d\vartheta }{d\rho }=\frac{1}{\rho ^{2}},
\end{equation*}%
so that%
\begin{equation*}
\vartheta =c_{1}-\rho ^{-1}.
\end{equation*}

\section{Convertible symmetry of order II}

Hidden symmetries have been defined with reference to the reduction of order
of an ODE \textit{by one}, and so far in the literature two types have been
studied. Specifically, a type I hidden symmetry is a Lie symmetry that is
lost when an ODE is reduced by one order, in addition to the symmetry used
to perform the reduction\cite{Barbara2}. Type II hidden symmetry appears in
addition to the inherited ones, when the ODE is reduced by one order. The
origin of hidden symmetries of type II has been attributed to reductions,
among others, that involve solvable, nonabelian, three-parameter groups and
which are associated with nonlocal symmetries\cite{Barbara2},\cite{Barbara3}%
. Non-local symmetries that reduce to point symmetries under a single
reduction of order have been labeled as "first-order" and termed "useful"
nonlocal symmetries\cite{Nonlocal}. Here we present an example which
introduces a \textit{useful, second-order nonlocal} symmetry in the
reduction of a fourth order ODE with a four-parameter solvable Lie algebra.
The term \textit{useful} should be understood here in a wider sense, since
this second-order nonlocal symmetry proves to be essential not only for the
integration of the initial fourth order equation, but also for solving the
third order ODE appearing after the first reduction in case it is an initial
equation to solve. In the reduction path, two of the symmetries are of
convertible type. One of the symmetries reappears as a point symmetry after
the second reduction of order (hence it is of order I) and the other becomes
a point symmetry after the third and final reduction (convertible symmetry
of order II).

In the following exposition we adhere to the notation of reference\cite%
{Mubarak}, so that $A_{n,m}$ denotes a Lie algebra of dimension $n$ and
isomorphism type $m$. The algebra $A_{4,1}$ has a basis%
\begin{equation}
X=\partial _{y},\text{ \ \ \ \ }Y=x\partial _{y},\text{ \ \ \ \ }%
Z=x^{2}\partial _{y},\text{ \ \ \ \ }U=\partial _{x},  \label{3.1}
\end{equation}%
with nonzero commutator relations%
\begin{equation}
\lbrack Y,U]=-X,\text{ \ \ \ \ }[Z,U]=-2Y.  \label{3.2}
\end{equation}%
Upon using the property of nonuniqueness of canonical coordinates (\ref{10a}%
), its most general invariant equation is%
\begin{equation}
\bigtriangleup ^{4}:y^{(iv)}=F(y^{\prime \prime \prime }).  \label{3.3}
\end{equation}%
The reduction path of eq. (\ref{3.3}) that we choose for analysis is
schematically presented in the following diagram%
\begin{equation*}
\begin{tabular}{l}
\begin{tabular}{l}
\underline{$\bigtriangleup ^{n},n-$equation order \ \ \ \ algebra; nonlocal
symmetries} \\ 
\\ 
$\bigtriangleup ^{4}$ $\ \ \ \ \ \ \ \ \ \ \ \ \ \ \ \ \ \ \ \ \ \ \ \ \ \ \
\ \ \ \ \ \ \ \ \ A_{4,1}(X,Y,Z,U)$ \\ 
$\downarrow $ $U$ \\ 
$\bigtriangleup ^{3}\ \ \ \ \ \ \ \ \ \ \ \ \ \ \ \ \ \ \ \ \ \ \ \ \ \ \ \
\ \ \ \ \ \ \ \ \ A_{1}(\tilde{X}^{(1)});\tilde{Y}^{(1),N},\tilde{Z}^{(1),N}$
\\ 
$\downarrow $ $\tilde{X}^{(1)}$ \\ 
$\bigtriangleup ^{2}$ $\ \ \ \ \ \ \ \ \ \ \ \ \ \ \ \ \ \ \ \ \ \ \ \ \ \ \
\ \ \ \ \ \ \ \ \ A_{1}(\tilde{Y}^{(2)});\tilde{Z}^{(2),N}$ \\ 
$\downarrow $ $\tilde{Y}^{(2)}$ \\ 
$\bigtriangleup ^{1}$ $\ \ \ \ \ \ \ \ \ \ \ \ \ \ \ \ \ \ \ \ \ \ \ \ \ \ \
\ \ \ \ \ \ \ \ \ A_{1}(\tilde{Z}^{(3)})$ \\ 
\\ 
\end{tabular}
\\ \hline
\end{tabular}%
\end{equation*}%
where superscript $N$ denotes that the symmetry is nonlocal and tilde
denotes restriction of the inherited symmetry generator to corresponding
fundamental differential invariants.

Definitions of type I and type II hidden symmetries carry certain
ambiguousness in the sense that their classification depends on the equation
to which they pertain to. Hence it may happen that the same symmetry may be
of different type along the reduction path depending on the equation to
which it is related to. For example, in the above reduction path symmetry $Y 
$ is a hidden symmetry of type I for equation $\bigtriangleup ^{4}$, while
it is a hidden symmetry of type II for equation $\bigtriangleup ^{3}$. On
the other hand symmetry $Z$ is a hidden symmetry of type I for equation $%
\bigtriangleup ^{4}$ while it is a hidden symmetry of type II for $%
\bigtriangleup ^{2}.$ This relative classification is avoided in making use
of convertible type symmetries since then for equation $\bigtriangleup ^{4}$
symmetry $Y$ is convertible symmetry of order I while $Z$ is convertible
type symmetry of order II.

\subsection{Analysis of the reduction path}

Symmetry $U$ is chosen to start the reduction procedure, so that based on
the commutator relations (\ref{3.2}) the only inherited symmetry is $X$,
while $Y$ and $Z$ become non-local. In order to predict what happens next
with these symmetries, we analyze the commutators of two sets; the first one
consists of symmetries $U,$ $X$ and $Y$ (and it is a subalgebra) and the
other of symmetries $U,Z$\ and $X.$ For the first, comparison with
expression (\ref{100}) indicates that $\gamma =0,$ $\alpha =0$ and $\beta =0,
$ and that $Y$ becomes a point symmetry following the next reduction using $%
\tilde{X}^{(1)}$.. On the other hand the second set does not form a
subalgebra because of the commutator $[U,Z]=2Y.$ Hence, $Z$ remains a
non-local symmetry during the second reduction too. In order to prove that $%
\tilde{Z}^{(3)}$ is a point symmetry we use the property, proved in the
Appendix A, that the commutator of $Z$ with each of the symmetries used for
reduction up to that point, yields the sum of already used symmetries scaled
with appropriate coefficients. Since the commutators of $Z$ with $U,X$ and $Y
$ yield $-2Y$, $0$ and $0$ respectively ($Y$ has been used in the reduction
previously), $\tilde{Z}^{(3)}$ is a point symmetry.

\subsection{Exploiting algebra solvability and properties of hidden
symmetries}

The group generated by $U$ has fundamental differential invariants $r=y$ and 
$v=s^{\prime }$ where $s=x,$ so that $U$ corresponds to $X,$ $X$ corresponds
to $Y$ and $Y$ corresponds to $Z$ of Case 2 discussed in Section $1$. In a
straightforward manner, the equation corresponding to (\ref{3.3}) becomes:%
\begin{equation}
v^{\prime \prime \prime }+15\frac{v^{\prime 2}}{v^{2}}-10\frac{v^{\prime
}v^{\prime \prime }}{v}=v^{5}F(3\frac{v^{\prime 2}}{v^{5}}-\frac{v^{\prime
\prime }}{v^{4}}).  \label{3.4}
\end{equation}%
Restrictions of $X,Y$ and $Z$ to $r$ and $v$ yield%
\begin{equation*}
\begin{tabular}{ll}
$X^{(1)}r=1$ & $X^{(1)}v=0$ \\ 
$Y^{(1)}r=x$ & $Y^{(1)}v=-\frac{1}{y^{\prime 2}}$ \\ 
$Z^{(1)}r=x^{2}$ & $Z^{(1)}v=-\frac{2x}{y^{\prime 2}}$%
\end{tabular}%
\end{equation*}%
so that

\begin{equation*}
\begin{tabular}{l}
$\tilde{X}^{(1)}=\partial _{r},$ \\ 
$\tilde{Y}^{(1),N}=x\partial _{r}-v^{2}\partial _{v}=(\int vdr)\partial
_{r}-v^{2}\partial _{v},$ \\ 
$\tilde{Z}^{(1),N}=x^{2}\partial _{r}-2xv^{2}\partial _{v}=(\int
vdr)^{2}\partial _{r}-2xv^{2}\partial _{v}.$%
\end{tabular}%
\end{equation*}%
The next reduction uses $\tilde{X}^{(1)}$ since it is the only point
symmetry. Introducing new canonical coordinates $\vartheta =r$ and $\rho =v,$
and changing variables so that $v^{\prime }=1/\vartheta ^{\prime }$ and $%
w=\vartheta ^{\prime },$ one obtains 
\begin{equation}
w^{\prime \prime }-3\frac{w^{\prime 2}}{w}-15\frac{w^{2}}{\rho ^{2}}-10\frac{%
w^{\prime }}{\rho }=w^{4}\rho ^{5}F(\frac{3}{\rho ^{5}w^{2}}+\frac{w^{\prime
}}{\rho ^{4}w^{3}}).  \label{3.5}
\end{equation}%
The following restrictions are readily calculated%
\begin{equation*}
\begin{tabular}{ll}
$Y^{(2)}\rho =-\rho ^{2}$ & $Y^{(2)}w=-3\rho w\partial _{w}$ \\ 
$Z^{(2)}\rho =-2x\rho ^{2}$ & $Z^{(2)}v=6x\rho w+\rho ^{3}w^{2}$%
\end{tabular}%
\end{equation*}%
yielding%
\begin{equation*}
\begin{tabular}{l}
$\tilde{Y}^{(2)}=\rho ^{2}\partial _{\rho }+3\rho w\partial _{w},$ \\ 
$\tilde{Z}^{(2),N}=-2\rho ^{2}\partial _{\rho }+(6x\rho w+\rho
^{3}w^{2})\partial _{w}.$%
\end{tabular}%
\end{equation*}%
Hence $Y$ generates point transformations of variables $\rho $ and $w,$
while $Z$ remains hidden (non-local). Focusing attention on the subalgebra $%
Span$ $(U,X$ and $Y)$ and the form $\tilde{Y}^{(2)}$ acquires as a point
symmetry, it is immediately apparent that $\tilde{Y}^{(2)}$ corresponds to $%
Z^{(2)}$ of Case $2$ (Section $1$), with $c_{1}=c_{4}=1$ as the only
coefficients different from zero. Continuing further along the reduction
path using the only point symmetry $\tilde{Y}^{(2)}$ and introducing new
canonical coordinates $t$ and $u$, the following relations are obtained
between new and old canonical coordinates:%
\begin{eqnarray*}
\rho  &=&-\frac{1}{t}, \\
w &=&-ut^{3},
\end{eqnarray*}%
so that eq. (\ref{3.5}) becomes%
\begin{equation*}
t^{\prime \prime }+\frac{3}{u}t^{\prime }=u^{4}F(u^{3}t^{\prime }).
\end{equation*}%
This second-order ODE may be reduced to a first-order equation by
substituting $\xi =t^{\prime },$ $\zeta =u,$ and since 
\begin{eqnarray*}
t &=&-\frac{1}{\rho }=-y^{\prime }, \\
u &=&\rho ^{3}w=-\frac{1}{y^{^{\prime \prime }}},
\end{eqnarray*}%
the following restrictions are obtained%
\begin{eqnarray*}
\tilde{Z}^{(3)}\zeta  &=&-\zeta ^{2}, \\
\tilde{Z}^{(3)}\xi  &=&-3\zeta \xi ,
\end{eqnarray*}%
so that finally $\tilde{Z}^{(3)}$ becomes a point symmetry 
\begin{equation}
\tilde{Z}^{(3)}=\zeta ^{2}\partial _{\zeta }+3\zeta \xi \partial _{\xi }.
\label{3.22}
\end{equation}

\section{Conclusion}

Convertible symmetries belonging to a class of hidden symmetries were
introduced. A convertible symmetry of order $n-1$ is defined as point
symmetry that disappears during the first reduction, remains hidden
(non-local) during $n-1$ reductions, and reappears as a point symmetry after 
$n$ reductions. In particular, properties of convertible symmetry of type I
(hidden symmetry that disappears after the first reduction of order of an $n$%
-th order ODE and reappears as a point symmetry after the second reduction)
were analyzed. It was shown that hidden symmetry, $Z$, along with two
symmetries, $X$ and $Y$, used for reduction necessarily form a subalgebra,
while $X$ and $Y$ necessarily commute. The convertible symmetry consists of
two parts one of which always remains the symmetry of point type. The other
part is the carrier of the main characteristics of the hidden symmetry and
determines the symmetry type along the reduction path. The importance of
convertible symmetry of type II (the symmetry that remains non-local during
two reductions of order and reappears as a point symmetry after the third
reduction) is illustrated by the reduction procedure in solving a fourth
order ODE, supplemented with the application of a practical rule for
determining the symmetry type along the reduction path.

\bigskip

{\Huge Appendices}

\appendix{}

\section{Determining symmetry type along the reduction path}

We address the problem of determining whether symmetry is of point or
nonlocal type along the reduction path. Essentially, the procedure is very
similar to the stepwise integration of the $n$-th order ODE having an $n$%
-dimensional Lie algebra, for example given in Section 5.3 of ref. \cite%
{Hydon}. Hence, the starting assumption is that the ODE of order $n$ has an $%
n$-dimensional solvable Lie algebra $L$. Suppose that the symmetry $\tilde{Z}
$ which appears after $k$ reductions, performed using generators $%
X_{1},.....X_{k},$ is of point type, i.e. its form is%
\begin{equation}
\tilde{Z}=\alpha (r_{k},v_{k})\partial _{r_{k}}+\beta (r_{k},v_{k})\partial
_{v_{k}}.  \label{a1}
\end{equation}%
Hence $Z$ is essentially equal to $X_{k+1}$ generator. Furthermore, suppose
that the generators $X_{1},.....X_{k}$ form a subalgebra of $L,$ so that $%
(r_{k},v_{k})$ are fundamental differential invariants of this subalgebra.
Symmetry $\tilde{Z}$ is therefore of the form (\ref{a1}) if $\tilde{Z}$ acts
on $(r_{k},v_{k})$ as a generator of point transformations, for at least one
of the functions $\alpha $ or $\beta $ different from zero. The property of
differential invariants is%
\begin{equation*}
\tilde{X}_{i}r_{k}=0,\text{ \ \ \ \ \ \ \ \ }\tilde{X}_{i}v_{k}=0,\text{\ \
\ \ \ \ \ \ \ \ \ }\forall i=1,...,k,
\end{equation*}%
so that%
\begin{eqnarray*}
\lbrack \tilde{X}_{i},\tilde{Z}]r_{k} &=&\tilde{X}_{i}\alpha , \\
\lbrack \tilde{X}_{i},\tilde{Z}]r_{k} &=&\tilde{X}_{i}\beta .
\end{eqnarray*}%
This can be written as%
\begin{eqnarray*}
(a_{ik+1}^{k-1}\tilde{X}_{k-1}+b\tilde{Z})r_{k} &=&b\alpha , \\
(a_{ik+1}^{k-1}\tilde{X}_{k-1}+b\tilde{Z})v_{k} &=&b\beta ,
\end{eqnarray*}%
where $a^{\prime }$s and $b$ are structure constants. From obvious
relationships%
\begin{eqnarray*}
b\alpha &=&\tilde{X}_{i}\alpha =0 \\
b\beta &=&\tilde{X}_{i}\beta =0,
\end{eqnarray*}%
follows that 
\begin{equation*}
b=0.
\end{equation*}%
Moreover, using property\cite{Olver}%
\begin{equation*}
\lbrack \tilde{X}_{i},\tilde{Z}]=\widetilde{[X_{i},Z]},
\end{equation*}%
we get the desired result, namely that the commutator of $Z$ with each of
the symmetries used for reduction up to that point, must yield the sum of
already used symmetries scaled with appropriate coefficients:

\begin{equation*}
\lbrack X_{i},Z]=c_{ik+1}^{j}X_{j},\text{ \ \ \ \ \ \ \ \ \ }1\leq j\leq k.
\end{equation*}

\section{Alternative Derivation of Convertible Symmetry Properties}

The reduction of the $n$-th order equation $\bigtriangleup ^{n}$ possessing
symmetries $X,$ $Y$ and $Z$ from the complete solvable algebra of
symmetries, is presented in the following diagram 
\begin{equation*}
\begin{tabular}{l}
$\bigtriangleup ^{n}$ $\ \ \ \ \ \ \ \ \ \ \ \ \ \ \ \ \ \ \ \ \ \ \ \ \ \
L(X,Y,Z,......)$ \\ 
$\downarrow $ $X$ \\ 
$\bigtriangleup ^{n-1}\ \ \ \ \ \ \ \ \ \ \ \ \ \ \ \ \ \ \ \ \ \ \ \
L^{(1)}(\tilde{Y}^{(1)},...),\tilde{Z}^{(1),N},...$ \\ 
$\downarrow $ $\tilde{Y}$ \\ 
$\bigtriangleup ^{n-2}$ $\ \ \ \ \ \ \ \ \ \ \ \ \ \ \ \ \ \ \ \ \ \ \
L^{(2)}(\tilde{Z}^{(2)},...),...$%
\end{tabular}%
.
\end{equation*}%
We adopt the convention that $X=X_{1},Y=X_{2}$ and $Z=X_{3}$. Following the
first reduction using vector field $X$, we assume that $Y$ remains a point
symmetry so that the condition 
\begin{equation}
\lbrack X,Y]=\lambda X,  \label{b1}
\end{equation}%
where $\lambda $ equals zero or it may be scaled to $1$, must be satisfied,
i.e., $X$ and $Y$ must form a $2$-dimensional subalgebra. In order for $Z$
to reappear as a point symmetry following reductions by $X$ and $\tilde{Y}$,
we again use the result proved in the Appendix A, namely that%
\begin{equation}
\lbrack Y,Z]=cX+dY.  \label{b2}
\end{equation}%
Since only symmetries $X$ and $Y$ have been used previously, there are no
additional terms on the right hand side of the above equation. For the
non-local symmetry $\tilde{Z}^{(1),N}$ the following condition has to be
satisfied

\begin{equation}
\lbrack X,Z]=bX+Y.  \label{b3}
\end{equation}%
where we have scaled the coefficient of $Y$ to $1$. In this case also we
have used the property proved in the Appendix A. Hence, we obtain the
following conditions upon symmetries $X,$ $Y$ and $Z$, 
\begin{eqnarray*}
\lbrack X,Y] &=&\lambda X, \\
\lbrack X,Z] &=&bX+X_{k}, \\
\lbrack Y,Z] &=&cX+dY.
\end{eqnarray*}%
\thinspace in addition to the Jacobi identity which holds for $X$, $Y$ and $%
Z $:%
\begin{equation*}
\lbrack X,[Y,Z]]+[Y,[Z,X]]+[Z,[X,Y]]=0.
\end{equation*}%
A straightforward calculation of the Jacobi identity yields 
\begin{equation*}
\lambda dX-\lambda Y=0
\end{equation*}%
so assuming $dX\neq Y$ we obtain%
\begin{equation*}
\lambda =0.
\end{equation*}%
Therefore, in order for the symmetry $Z$ to disappear during the first
reduction and reappear following the second reduction it is necessary that
following conditions be fulfilled:

\begin{eqnarray*}
\lbrack X,Y] &=&0, \\
\lbrack X,Z] &=&bX+Y, \\
\lbrack Y,Z] &=&cX+dY.
\end{eqnarray*}

\end{document}